\documentclass[a4paper,10pt,twoside]{cpc-hepnp}

\usepackage{multicol}
\usepackage{booktabs}
\usepackage{amssymb,bm,mathrsfs,bbm,amscd}
\usepackage[tbtags]{amsmath}
\usepackage{lastpage}
\usepackage[dvips]{graphicx,color}

\begin{document}

\fancyhead[co]{\footnotesize SHEN Cheng-Ping: $XYZ$ particles at
Belle}

\footnotetext[0]{Received ??? 2009}

\title{$XYZ$ particles at Belle\thanks{Supported by the National Natural Science Foundation of China under
Contract Nos. 10825524 and 10935008 and the Department of Energy
under Contract No. DE-FG02-04ER41291 (U Hawaii).}}

\author{%
SHEN Cheng-Ping$^{1)}$ \\(for the Belle Collaboration)\email{shencp@phys.hawaii.edu}%
} \maketitle

\address{%
1~(University of Hawaii, Honolulu, HI 96822, USA)\\
}

\begin{abstract}
In this paper, I review recent progress in the study of the $XYZ$
particles at Belle. I only focus on studies with charmonium and one
or more light mesons in the final states. This covers the $X(3872)$,
$X(3915)$, $Y(4140)$, $X(4350)$, and the charged $Z$ states.
\end{abstract}



\begin{multicols}{2}

\section{Introduction}

Recent experimental observations near the charm threshold strongly
suggest that the spectrum of resonances with hidden charm is
remarkably more rich than suggested by the standard
quark-antiquark template and very likely includes states where the
heavy-quark $c\bar{c}$ pair is accompanied by light quarks and/or
gluons. Lots of new charmonium-like resonances ($XYZ$ particles)
in the B factories have been observed in the final states with a
charmonium and some light hadrons. They could be candidates for
usual charmonium states, however, there are also lots of strange
properties shown from these states.

The following $XYZ$ particles, that I will consider in this paper,
are the $X(3872)$,  $X(3915)$,  $Y(4140)$, $X(4350)$, and the
charged $Z$ states. The $X(3915)$ and $X(4350)$ found in two-photon
processes, and the $Y(4140)$ found in $B$ decays are new
observations. The results of searching for possible $X/Y$ states in
the $\Upsilon(1S)$ radiative decays are also reported here for the
first time.

\section{The {\boldmath $X(3872)$}}

The $X(3872)$ was discovered by Belle in 2003~\cite{belle_x3872}
as a narrow peak in the $\pi^+\pi^-J/\psi$ invariant mass
distribution from $B\to K \pi^+\pi^-J/\psi$ decays. This discovery
mode was remeasured with more statistics at Belle. Belle reported
a new result for the mass of the $X(3872)$ as $M^{Belle}_{X(3872)}
= 3871.46\pm 0.37\pm 0.07$~MeV~\cite{belle_x3872_mass}. The most
precise measurement of the mass was reported by CDF using the same
decay channel: $M^{CDF}_{X(3872)} = 3871.61\pm 0.16\pm
0.19$~MeV~\cite{CDF_x3872_mass}. A new world average that includes
these new measurements plus other results that use the $\pi^+\pi^-
J/\psi$ decay mode is $M^{\rm avg}_{X(3872)} = 3871.46\pm
0.19$~MeV, which is very close to the $D^{*0}\bar{D^0}$ mass
threshold: $m_{D^{*0}}+m_{D^0} = 3871.81 \pm 0.36$~MeV~\cite{PDG}.
This suggests a binding energy of $-0.35\pm 0.41$~MeV if $X(3872)$
is interpreted as a $D^{*0}\bar{D^0}$ molecule. Belle also
reported the first statistically significant observation of $B^0
\to X(3872) K^0_S$ and measured the ratio of branching fractions
to be
\[
\frac{{\cal B}(B^0 \to X(3872) K^0)}{ {\cal B}(B^+ \to X(3872)K^+)}
= 0.82\pm0.22\pm0.05,
\]
consistent with unity. The mass difference between the $X(3872)$
states produced in $B^+$ and $B^0$ decay is found to be
$M^{B^+}_{X} - M^{B^0}_{X} = 0.18\pm 0.89\pm 0.26$~MeV, consistent
with zero.

In addition, Belle did a study of $X(3872)$ production in
association with a $K\pi$ in $B^0\to K^+\pi^- \pi^+\pi^-J/\psi$
decays~\cite{belle_x3872_mass}. In a sample of 657M $B\bar{B}$
pairs a signal of about 90 $X(3872)\to \pi^+\pi^- J/\psi$ events
was observed. Unlike the $B^0\to K^+\pi^- +{charmonium}$ where
$K^+\pi^-$ is mainly from $K^*(892)$ decays, it is evident that
most of the $K\pi$ pairs have a phase space-like distribution,
with little or no signal for $K^*(892)\to K\pi$. Belle measures
${\cal B}(B^0 \to X(3872)(K^+ \pi^-)_{NR}){\cal B}(X(3872)\to
\pi^+\pi^- J/\psi)=(8.1\pm2.0^{+1.1}_{-1.4})\times 10^{-6}$ and
sets the 90\% C.L. limit, ${\cal B}(B^0 \to X(3872) K^*(892)){\cal
B}(X(3872)\to \pi^+\pi^- J/\psi) <3.4\times 10^{-6}$. Belle
reports a $K^*(892)$ to $K\pi$ non-resonant ratio of
\[
\frac{{\cal B}(B\to (K^+\pi^-)_{K^*(892)}J/\psi)}{{\cal B}(B\to
(K^+\pi^-)_{NR}J/\psi)}<0.55,
\]
at the 90\% C.L.~\cite{olsen}. This is an indication that the
$X(3872)$ state is not a conventional charmonium state. However,
there is no solid calculation of the above ratio assuming different
nature for the $X(3872)$ state.

BaBar studied $B\to K D^{*0}\bar{D^0}$ with a sample of 383M
$B\bar{B}$ pairs and found a similar near-threshold enhancement
that, if considered to be due to the $X(3872)\to D^{*0}\bar{D^0}$,
gave a mass of $3875.1^{+0.7}_{-0.5}\pm
0.5$~MeV~\cite{babar_x3872_ddstr}. This state has been considered to
be a state different from the $X(3872)$ in the literature. However,
a subsequent Belle study of $B\to KD^{*0}\bar{D^0}$ based on 657M
$B\bar{B}$ pairs was performed for both $D^{*0} \to D^0 \gamma$ and
$D^{*0} \to D^0 \pi^0$ decay modes. Belle found a signal of
$50.1^{+14.8}_{-11.1}$ events with a mass of
$3872.9^{+0.6+0.4}_{-0.4-0.5}$~MeV, a width of
$3.9^{+2.8+0.2}_{-1.4-1.1}$ MeV by fitting the peak on the
$D^{*0}\bar{D^0}$ invariant mass distribution with a phase-space
modulated Breit-Wigner (BW) function~\cite{belle_ddstar}. The
branching fraction of ${\cal B}(B \to X(3872) K){\cal B}(X(3872)\to
D^{*0}\bar{D^0})$ was measured to be $(0.80\pm0.20\pm0.10)\times
10^{-4}$. The significance of the signal is 6.4$\sigma$. The
difference between the X(3872) mass and the $D^{*0}\bar{D^0}$
threshold is calculated to be $1.1^{+0.6+0.1}_{-0.4-0.3}$~MeV.

\section{The {\boldmath $X(3915)$}}

Three  new (neutral) states have been discovered by Belle in the
3.90-3.95 GeV region. The $X(3940)$ has been found in the double
charmonium production process, and its sizable decay to $D^*\bar{D}$
is confirmed~\cite{belle_x3940}. The Y (3940) has been observed in
the B decay process $B^-\to Y(3940) K^-$ and $Y(3940) \to \omega
J/\psi$~\cite{belle_y3940}, and is a candidate for an exotic state,
such as a hybrid meson ($c \bar{c} g$). The Z(3930) has been found
as a $D\bar{D}$ mass peak in $\gamma\gamma\to D\bar{D}$
events~\cite{belle_z3930}, and is usually assigned to a $2^3P_2$
$c\bar{c}$ charmonium state, which is commonly called the
$\chi_{c2}^{\prime}$. These three states appear in different
production and decay processes, and are usually considered to be
distinct particles. However there is no decisive evidence for this.
Predictions of partial decay widths of $Y(3940)$ to $\gamma \gamma$
and $\omega J/\psi$ states was calculated based on an interpretation
of $Y(3940)$ as a $D^*\bar{D}^*$ bound state~\cite{tanja}.

To add more information on the states in this mass region, it is
important to search for a signature of $Y(3940)$ or any other
resonant state decaying to $\omega J/\psi$ in two-photon processes.
This final state is the lightest combination of two vector mesons
with definite C-even and $I=0$ quantum numbers that can be produced
in two-photon processes via a hidden-charm state. This analysis is
based on a 694~fb$^{-1}$ data sample collected at the
$\Upsilon(nS)~(n=3,4,5)$ resonances and 60 MeV below the
$\Upsilon(4S)$ resonance.

Belle observed a dramatic and rather narrow peak around 3.92 GeV,
$X(3915)$, in the cross section for $\gamma\gamma\to \omega
J/\psi$~\cite{belle_y3915}. The invariant mass distribution for
the $\omega J/\psi$ candidates produced in $\gamma\gamma$
collision, shown in Fig.~\ref{fig:omegajpsi}, shows a sharp peak
near threshold and not much else. It is far above the non-$\omega
J/\psi$ background contribution which is estimated by the events
in the $\omega$ and $J/\psi$ mass sidebands (shown shaded for
comparison). An unbinned maximum likelihood fit was performed to
the 73 events in the region from 3.875 GeV to 4.2 GeV.

\begin{center}
\includegraphics[width=8cm]{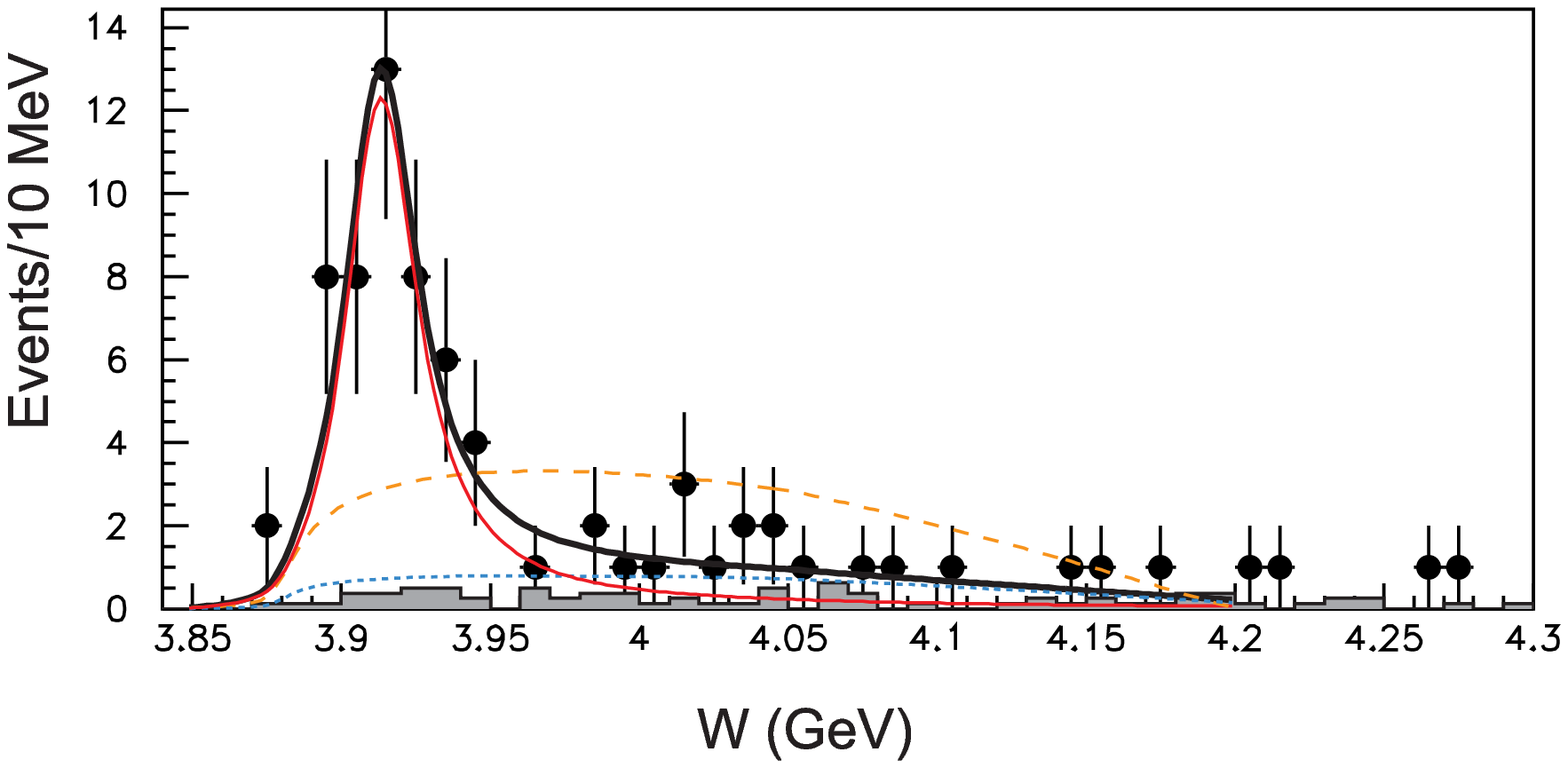}
\figcaption{The $\omega J/\psi$ mass distribution for selected
events in the $\gamma \gamma \to \omega J/\psi$ process. The dots
with error bars are the experimental data. The shaded histogram is
the distribution of non-$\omega J/\psi$ backgrounds estimated by the
sideband distributions. The bold solid, thinner solid and the lower
dashed curves are the total, resonance and background contributions,
respectively, from the fit. The upper dot-dashed curve is the fit
without assuming a resonance.} \label{fig:omegajpsi}
\end{center}

A fit with an S-wave BW function with a variable width for the
resonance component plus a smooth background function gives results
for the resonance parameters of the $X(3915)$: $M = (3915\pm 3\pm
2)~{\rm MeV},~ \Gamma = (17\pm 10 \pm 3)~{\rm MeV}$. The statistical
significance of the signal is $7.7\sigma$. This value for the mass
is about $2\sigma$ different from that of the $Z(3930)$ ($M=3929\pm
5\pm 2$~MeV), indicating that these two peaks may not be different
decay channels of the same state. On the other hand, there is good
agreement between these preliminary results and the mass and width
quoted by BaBar for the $Y(3940)$, which is also seen in $\omega
J/\psi$.

The product of the $X(3915)$ two-photon decay width and the
branching fraction to $\omega J/\psi$ depends on the $J^{P}$
value. Belle determines
\[
\Gamma_{\gamma\gamma}(X(3915)){\cal B}(X(3915)\to \omega J/\psi)
=61\pm 17 \pm 8~{\rm eV},
\]
or
\[
\Gamma_{\gamma\gamma}(X(3915)){\cal B}(X(3915)\to \omega J/\psi)
=18\pm 5 \pm 2~{\rm eV},
\]
for $J^{P}=0^{+}$ or $2^{+}$, respectively.

Based on this result, and the measured width $\Gamma$, the product
of the two partial widths of the $X(3915)$, $\Gamma_{\gamma
\gamma}(X)\Gamma_{\omega J/\psi}(X)$ is of order $10^3$ keV$^2$. If
we assume $\Gamma_{\gamma \gamma} \sim \mathcal{O}$ (1 keV), typical
for an excited charmonium state, this implies $\Gamma_{\omega
J/\psi}\sim \mathcal{O}$ (1 MeV), a rather large value, even for the
charmonium-inclusive partial width of such a state. Predictions of
the partial decay widths of $Y(3940)$ based on a $D^*\bar{D}^*$
bound-state model~\cite{tanja} obtains a  product roughly compatible
to the present measurement.

\section{The {\boldmath $Y(4140)$} and {\boldmath $X(4350)$}}

Using exclusive $B^+ \to J/\psi \phi K^+$ decays, the CDF
Collaboration observed a narrow structure near the $J/\psi \phi$
mass threshold ($m_{J/\psi}+m_{\phi}=4.117~\hbox{GeV}/c^2$) with a
statistical significance of 3.8$\sigma$~\cite{CDF}. The mass and
width of this structure are fitted to be $4143.0\pm 2.9\pm
1.2~\hbox{MeV}$ and $11.7^{+8.3}_{-5.0}\pm 3.7~\hbox{MeV}$,
respectively using an $S$-wave relativistic BW function. Assuming
isospin conservation, this new state, called $Y(4140)$ by the CDF
Collaboration, is an isospin singlet state with positive $C$ and $G$
parities since the quantum numbers of both $J/\psi$ and $\phi$ are
$I^G(J^{PC})=0^-(1^{--})$. It was argued by the CDF Collaboration
that the $Y(4140)$ can not be a conventional charmonium state,
because a charmonium state with mass about 4143~$\hbox{MeV}$ would
dominantly decay into open charm pairs, and the branching fraction
into the doubly OZI (Okubo-Zweig-Iizuka) forbidden modes $J/\psi
\phi$ or $J/\psi \omega$ would be negligible.

There have been a number of different interpretations proposed for
the $Y(4140)$, including a $D_{s}^{\ast+} {D}_{s}^{\ast-}$
molecule~\cite{tanja, liux, ding, namit,liu3,huang, raphael,molina},
an exotic $1^{-+}$ charmonium hybrid~\cite{namit}, a
$c\bar{c}s\bar{s}$ tetraquark state~\cite{stancu}, or a natural
consequence of the opening of the $\phi J/\psi$ channel~\cite{eef}.
There are also arguments that the $Y(4140)$ should not be a
conventional charmonium $\chi_{c0}^{\prime\prime}$ or
$\chi_{c1}^{\prime\prime}$~\cite{liu2}, nor a scalar $D_{s}^{\ast+}
{D}_{s}^{\ast-}$ molecule since QCD sum rules~\cite{wangzg, wangzg2}
predict masses inconsistent with the observed mass.

The Belle Collaboration searched for this state using the same
process with $772\times 10^6$ $B \bar{B}$ pairs. Figure~\ref{my4140}
shows the $\phi J/\psi$ invariant mass distribution with the energy
difference $\Delta E=E_B-E_{beam}$ restricted to the range $|\Delta
E|<40$ MeV, where $E_B$ is the center-of-mass (CM) energy of the B
candidate and $E_{beam}$ is the CM beam energy. No significant
$Y(4140)$ signal was found. From the fit with the mass and width of
the $Y(4140)$ fixed to the CDF  measurements, we obtain
$7.5^{+4.9}_{-4.4}$ signal events. The statistical significance of
the $Y(4140)$ is estimated to be $1.9\sigma$ and the upper limit on
the production rate ${\cal B}(B^+\to Y(4140)K^+){\cal B} (Y(4140)\to
J/\psi \phi)$ is measured to be $6\times 10^{-6}$ at the 90\% C.L.
Although this upper limit is lower than the central value of the CDF
measurement $(9.0\pm 3.4\pm 2.9)\times 10^{-6}$, it does not
contradict the CDF measurement considering the large
error~\cite{CDF}. \vspace{0.8cm}

\begin{center}
\includegraphics[width=7cm]{fig2.epsi}
\end{center}
\figcaption{\label{my4140}Fit to the $\phi J/\psi$ invariant mass
distribution with the mass and width of the $Y(4140)$ fixed to the
CDF measurements within $|\Delta E|<40$ MeV region.}

Assuming the $Y(4140)$ is a $D_{s}^{\ast+} {D}_{s}^{\ast-}$ molecule
with quantum number $J^{PC}=0^{++}$ or $2^{++}$, the authors of
Ref.~\cite{tanja} predicted a two-photon partial width of the
$Y(4140)$ of the order of 1~keV, which is large and can be tested
with experimental data. The Belle Collaboration searched for this
state in two-photon production~\cite{x4350} to test this model. This
analysis is based on a 825~fb$^{-1}$ data sample collected at the
$\Upsilon(nS)~(n=1,3,4,5)$ resonances. No $Y(4140)$ signal is
observed, and the upper limit on the product of the two-photon decay
width and branching fraction of $Y(4140) \to \phi J/\psi$ is
measured to be $\Gamma_{\gamma \gamma}(Y(4140)) {\cal
B}(Y(4140)\to\phi J/\psi)<39~\hbox{eV}$ for $J^P=0^+$, or
$<5.7~\hbox{eV}$ for $J^P=2^+$ at the 90\% C.L. for the first time.
The upper limit on $\Gamma_{\gamma \gamma}(Y(4140)) {\cal
B}(Y(4140)\to\phi J/\psi)$ from this experiment is lower than the
prediction of $176^{+137}_{-93}$~eV for $J^{PC}=0^{++}$, or
$189^{+147}_{-100}$~eV for $J^{PC}=2^{++}$ (calculated by using the
numbers in Ref.~\cite{tanja} and total width of the $Y(4140)$ from
CDF measurement~\cite{CDF}). This disfavors the scenario of the
$Y(4140)$ being a $D_{s}^{\ast+} {D}_{s}^{\ast-}$ molecule with
$J^{PC}=0^{++}$ or $2^{++}$.

Evidence is reported for a narrow structure at $4.35~\hbox{GeV}/c^2$
in the $\phi J/\psi$ mass spectrum in the above two-photon process
$\gamma \gamma \to \phi J/\psi$ (see Fig.~\ref{mkkjpsi-fit2}) in the
Belle experiment. In order to obtain the resonance parameters of the
structure at $4.35~\hbox{GeV}/c^2$, an unbinned extended maximum
likelihood method is applied to the $\phi J/\psi$ mass spectrum. The
distribution is fitted in the range 4.2 to 5.0~GeV/$c^2$ with an
acceptance-corrected BW function convoluted with a double Gaussian
resolution function as the signal shape and a constant term as the
background shape. The shape of the double Gaussian resolution
function is obtained from MC simulation.

\begin{center}
\includegraphics[width=8cm]{fig3.epsi}
\end{center}
\figcaption{\label{mkkjpsi-fit2} The $\phi J/\psi$ invariant mass
distribution of the final candidate events in the $\gamma \gamma \to
\phi J/\psi$ process. The blank histogram is the experimental data.
The fit to the $\phi J/\psi$ invariant mass distribution from 4.2 to
5.0 GeV/$c^2$ is described in the text. The solid line is the best
fit, the dashed line is the background, and the shaded histogram is
from normalized $\phi$ and $J/\psi$ mass sidebands. The arrow shows
the position of the $Y(4140)$.}

From the fit, a signal of $8.8^{+4.2}_{-3.2}$ events, with
statistical significance of 3.2 standard deviations including
systematic uncertainty, is observed. The mass and natural width of
the structure (named $X(4350)$) are measured to be
$4350.6^{+4.6}_{-5.1}\pm 0.7~\hbox{MeV}$ and $13.3^{+17.9}_{-9.1}\pm
4.1~\hbox{MeV}$, respectively. The products of its two-photon decay
width and branching fraction to $\phi J/\psi$ is measured to be
$\Gamma_{\gamma \gamma}(X(4350)) B(X(4350)\to\phi
J/\psi)=6.4^{+3.1}_{-2.3}\pm 1.1~\hbox{eV}$ for $J^P=0^+$, or
$1.5^{+0.7}_{-0.5}\pm 0.3~\hbox{eV}$ for $J^P=2^+$. It is noted that
the mass of this structure is consistent with the predicted values
of a $c\bar{c}s\bar{s}$ tetraquark state with $J^{PC}=2^{++}$ in
Ref.~\cite{stancu} and a $D^{\ast+}_s {D}^{\ast-}_{s0}$ molecular
state in Ref.~\cite{zhangjr}.

\section{The charged {\boldmath $Z$} states}

Belle's $Z(4430)^+$ signal is a sharp peak in the $\pi^+\psi(2S)$
invariant mass distribution from $B\to K\pi^+\psi(2S)$
decays~\cite{belle_z4430}. A fit using a BW function gives
$M=4433\pm 4\pm 2$~MeV and $\Gamma = 45^{+18+30}_{-13-13}$~MeV, with
an estimated statistical significance of more than $6\sigma$.
Consistent signals are seen in various subsets of the data: {\it
i.e.} for both the $\psi(2S)\to \ell^+\ell^-$ and $\psi(2S)\to
\pi^+\pi^- J/\psi$ subsamples, the $\psi(2S)$($J/\psi$)$\to e^+e^-$
and $\mu^+\mu^-$ subsamples, etc.

However the BaBar group did not confirm the $Z(4430)^+\rightarrow
\pi^{+}\psi(2S)$ mass peak in their partial wave analysis of $B\to
K\pi\psi(2S)$ decays~\cite{babar_z4430} although statistically BaBar
result does not contradict Belle's observation. Belle performed a
reanalysis of their data with a similar partial wave analysis.
Specifically, they modelled  $B\to K\pi\psi(2S)$ as the sum of
two-body decays $B\to K_i^*\psi(2S)$, where $K_i^*$ denotes all of
the known $K^*\to K\pi$ resonances that are kinematically
accessible, both with and without a $B\to K Z$ component, where $Z$
denotes a resonance that decays to
$\pi\psi(2S)$~\cite{belle_z4430_dalitz}.

The data points in Fig.~\ref{fig:z4430_dalitz-analysis} shows the
$M^2(\pi\psi(2S))$ Dalitz plot projection with the prominent $K^*$
bands removed compared with the results of the fit with no $Z$
resonance, shown as a dashed histogram, and that with a $Z$
resonance, shown as the solid histogram. The fit with the $Z$ is
favored over the fit with no $Z$ by $6.4\sigma$.  The fitted mass,
$M=4443^{+15+19}_{-12-13}$~MeV, agrees within the systematic
errors with the earlier Belle result; the fitted width, $\Gamma =
107^{+86+74}_{-43-56}$~MeV, is larger, but also within the
systematic errors of the previous result. The product branching
fraction from the Dalitz fit: ${\cal B}(B^0\to  K Z^+) {\cal
B}(Z^+ \to \pi^+\psi(2S)) = (3.2^{+1.8+9.6}_{-0.9-1.6})\times
10^{-5}$ is not in strong contradiction with the BaBar 95\% C.L.
upper limit of $3.1\times 10^{-5}$.

\begin{center}
\includegraphics[width=7cm]{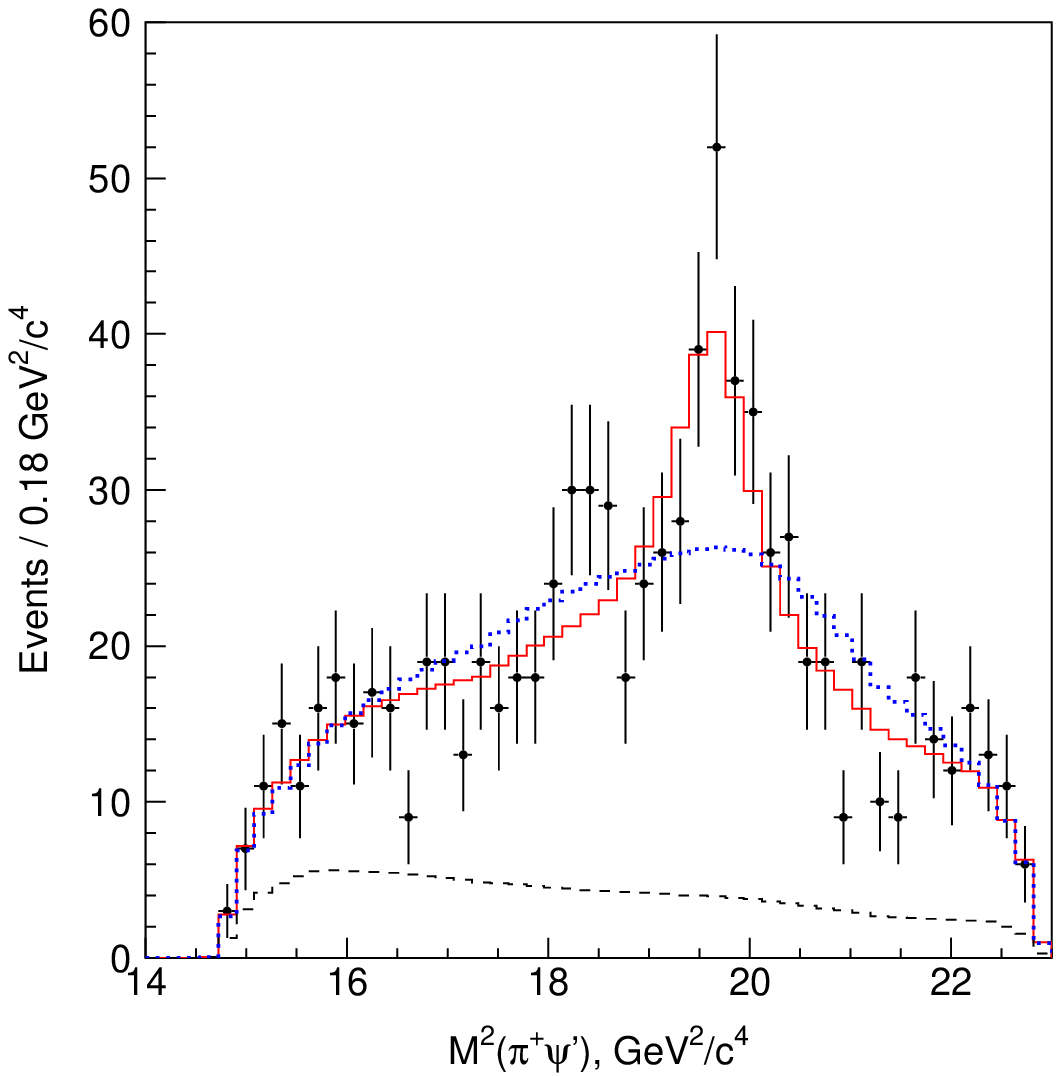}
\figcaption{The data points show the $M^2(\pi\psi(2S))$ projection
of the Dalitz plot with the $K^*$ bands removed from $B\to K \pi^+
\psi(2S)$ decays. The histograms show the corresponding projections
of the fits with and without a $Z\to \pi\psi(2S)$ resonance term.}
\label{fig:z4430_dalitz-analysis}
\end{center}

In addition to the $Z(4430)^+$, Belle has presented results of an
analysis of $B\to K\pi^+\chi_{c1}$ decays that require two
resonant states in the $\pi^+\chi_{c1}$
channel~\cite{belle_z14050}. In this case the kinematically
allowed mass range for the $K\pi$ system extends beyond the
$K^*_3(1780)$ $F$-wave resonance and $S$-, $P$-, $D$- and $F$-wave
terms for the $K\pi$ system are included in the model. The fit
with a single resonance in the $Z\to  \pi\chi_{c1}$ channel is
favored over a fit with only $K^*$ resonances and no $Z$ by more
than $10\sigma$.  Moreover, a fit with two resonances in the
$\pi\chi_{c1}$ channel is favored over the fit with only one $Z$
resonance by $5.7\sigma$. The fitted masses and widths of these
two resonances are: $M_1=4051\pm 14^{+20}_{-41}$~MeV and $\Gamma_1
= 82^{+21+47}_{-17-22}$~MeV and $M_2=4248^{+44+180}_{-29-35}$~MeV
and $\Gamma_2 = 177^{+54+316}_{-39-61}$~MeV. The product branching
fractions have central values similar to that for the $Z(4430)$
but with large errors. Figure~\ref{fig:z4050_dalitz-analysis}
shows the $M(\pi\chi_{c1})$ projection of the Dalitz plot with the
$K^*$ bands excluded and the results of the fit with no $Z\to
\pi\chi_{c1}$ resonances and with two $Z\to \pi\chi_{c1}$
resonances.

\begin{center}
\includegraphics[width=7cm]{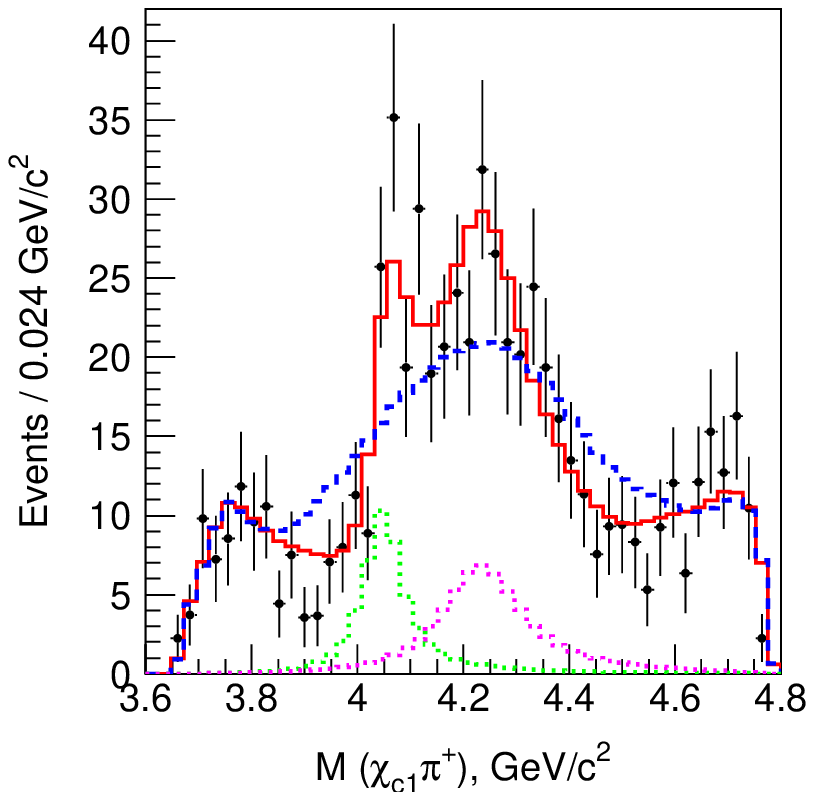}
\figcaption{The data points show the $M(\pi\chi_{c1})$ projection of
the Dalitz plot with the $K^*$ bands removed from $B\to K \pi^+
\chi_{c1}$ decays. The histograms show the corresponding projections
of the fits with and without the two $Z\to \pi\chi_{c1}$ resonance
terms.} \label{fig:z4050_dalitz-analysis}
\end{center}

Since the $Z$ states have hidden charm and light quarks to allow
them to decay to charmonium rich final states and with non-zero
charge, if any one of them is confirmed, it is an unambiguous
evidence for a state with more than three quarks.

\section{The X/Y states in {\boldmath $\Upsilon(1S)$} radiative decays}

For charge parity even $X/Y$ states, one way to study them is
through radiative decays of the $\Upsilon$ states below open-bottom
threshold. The production rates of the lowest lying $P$-wave
spin-triplet ($\chi_{cJ}$, $J$=0, 1, or 2) and $S$-wave spin-singlet
($\eta_c$) have been calculated in Ref.~\cite{ktchao}, and the
former is at a few millionth level while the latter is about
$5\times 10^{-5}$. There is no existing calculation of the excited
charmonium states, let alone the $X(3872)$, $Y(4140)$ and $X(3915)$.

The Belle Collaboration searched for the $X(3872)$, $Y(4140)$ and
$X(3915)$ in $\Upsilon(1S)$ radiative decays~\cite{y1s-decays} based
on 5.8 fb$^{-1}$ of data collected at the $\Upsilon(1S)$ and 1.8
fb$^{-1}$ of data collected at 9.43 GeV. In the $\pi^+ \pi^- J/\psi$
mode, except for the ISR produced $\psi(2S)$, there are only a few
events scatter above the $\psi(2S)$ peak (see
Fig.~\ref{mppjpsi-y1s}). There is only one event in the $X(3872)$
mass region. No $X(3872)$ nor $X(3915)$ signals was observed in the
$\pi^+\pi^-\pi^0 J/\psi$ mode (see Fig.~\ref{m3pijpsi-y1s}). Upper
limits on the production rates are determined to be ${\cal
B}(\Upsilon \to \gamma X(3872))\times{\cal
B}(X(3872)\to\pi^+\pi^-J/\psi)< 2.2 \times 10^{-6}$, ${\cal
B}(\Upsilon\to \gamma X(3872))\times{\cal
B}(X(3872)\to\pi^+\pi^-\pi^0 J/\psi)< 3.4\times 10^{-6}$ and ${\cal
B}(\Upsilon \to \gamma X(3915))\times{\cal B}(X(3915)\to\omega
J/\psi)< 3.4\times 10^{-6}$ at the 90\% C.L. respectively. Belle
also searched for the $Y(4140)$, but there is no candidate event in
the signal region, and the upper limit on the production rate ${\cal
B}(\Upsilon \to \gamma Y(4140)){\cal B}(Y(4140)\to\phi J/\psi))$ is
determined to be $2.6\times 10^{-6}$ at the 90\% C.L.

\begin{center}
\includegraphics[width=7cm]{fig6.epsi}
\figcaption{$\pi^+ \pi^- J/\psi$ invariant mass distribution of
the selected $\Upsilon(1S)\to \gamma \pi^+ \pi^- J/\psi$
candidates. Dots with error bars are data, blank histograms are MC
expectation of the signal shape with arbitrary normalization.}
\label{mppjpsi-y1s}
\end{center}

\begin{center}
\includegraphics[width=7cm]{fig7.epsi}
\figcaption{$\pi^+ \pi^- \pi^0 J/\psi$ invariant mass distribution
of the selected $\Upsilon(1S)\to \gamma \pi^+ \pi^- \pi^0 J/\psi$
candidates. Dots with error bars are data, blank histograms are MC
expectation of the signal shape with arbitrary normalization. }
\label{m3pijpsi-y1s}
\end{center}


\section{Summary}

In summary, there are lots of charmonium-like $XYZ$ states states
observed recently in charmonium mass region, but many of them show
properties different from the naive expectation of conventional
charmonium states. It has been suggested that many of the $XYZ$
states are multiquark states, either tetraquarks or molecules. The
problem with the tetraquark explanation is that it predicts
multiplets with other charge states that have not been observed,
and larger widths than have been observed. The possibility that
some of the $XYZ$ states are molecules is likely intertwined with
threshold effects that occur when channels are opened up.
Including coupled-channel effects and the rescattering of charmed
meson pairs in the mix can also result in shifts of the masses of
$c\bar{c}$ states and result in meson-meson binding which could
help explain the observed spectrum. However, due to limited
statistics, the experimental information on the properties of any
of these states is not enough for us to draw solid conclusion, let
alone our poor knowledge on the QCD prediction of the properties
of the exotic states or the usual charmonium states.

Many of the $XYZ$ states need independent confirmation and to
understand them will require detailed studies of their properties.
With better experimental and theoretical understanding of these
states we will have more confidence in believing that any of these
new states are non-conventional $c\bar{c}$ states like molecules,
tetraquarks, and hybrids. In the near future, the Belle II
experiment~\cite{belle2} under construction, with about
50~ab$^{-1}$ data accumulated, will surely improve our
understanding of all these states.

\acknowledgments{We borrowed many material from arXiv:0909.2713 by
Steve Olsen and arXiv:0910.3138 by  Chang-Zheng Yuan, where the
same topic was discussed based on the same experimental
information. We thank the organizers for their kind invitation and
congratulate them for a successful workshop.
}

\end{multicols}

\vspace{-2mm}
\centerline{\rule{80mm}{0.1pt}}
\vspace{2mm}

\begin{multicols}{2}

\end{multicols}

\vspace{5mm}

\clearpage


\begin{thebibliography}{90}

\vspace{3mm}
\bibitem{belle_x3872}  Choi~S~K {\em et al.} (Belle Collaboration),
Phys. Rev. Lett., 2003, {\bf 91}: 262001

\bibitem{belle_x3872_mass} Adachi~I {\em et al.} (Belle Collaboration),
arXiv:0809.1224.

\bibitem{CDF_x3872_mass}  Abulencia~A {\em et al.} (CDF Collaboration),
Phys. Rev. Lett., 2009, {\bf 103}: 152001

\bibitem{PDG} Amsler~C  {\em et al.} (Particle Data Group),
Phys. Lett. B, 2008, {\bf 667}: 1

\bibitem{olsen} Olsen~S~L, arXiv:0909.2713

\bibitem{babar_x3872_ddstr}  Aubert~B {\em et al.} (BaBar Collaboration),
Phys. Rev. D, 2008, {\bf 77}: 011102

\bibitem{belle_ddstar} Adachi~I {\em et al.} (Belle Collaboration),
arXiv:0810.0358

\bibitem{belle_x3940} Abe K  {\em et al.} (Belle Collaboration),
Phys. Rev. Lett., 2007, {\bf 98}: 082001

\bibitem{belle_y3940} Choi~S~K {\em et al.} (Belle Collaboration),
Phys. Rev. Lett., 2005, {\bf 94}: 182002

\bibitem{belle_z3930} Uehara S {\em et al.} (Belle Collaboration),
Phys. Rev. Lett., 2006, {\bf 96}: 082003

\bibitem{tanja} Branz T, Gutsche T, and Lyubovitskij V E,
Phys. Rev. D, 2009, {\bf 80}: 054019

\bibitem{belle_y3915} Uehara S {\em et al.} (Belle Collaboration),
in preparation.

\bibitem{CDF} Aaltonen~T {\em et al.} (The CDF Collaboration),
Phys. Rev. Lett., 2009, {\bf 102}: 242002

\bibitem{liux} Liu~X and Zhu~S~L, Phys. Rev. D, 2009, {\bf 80}: 017502

\bibitem{ding} Ding~G~J, arXiv:0904.1782

\bibitem{namit} Mahajan~N, arXiv:0903.3107

\bibitem{liu3} Liu~X and Ke~H~W, arXiv:0907.1349

\bibitem{huang} Zhang~J~R and Huang~M~Q, arXiv:0906.0090

\bibitem{raphael} Albuquerque~R~M, Bracco~M~E and Nielsen~M, Phys. Lett. B, 2009, {\bf 678}: 186

\bibitem{molina} Molina~R and Oset~E, arXiv:0907.3043

\bibitem{stancu} Stancu~Fl, arXiv:0906.2485

\bibitem{eef} Beveren E~van and Rupp~G, arXiv:0906.2278

\bibitem{liu2} Liu~X, Phys. Lett. B, 2009, {\bf 680}: 137

\bibitem{wangzg} Wang~Z~G, Eur. Phys. J. C, 2009, {\bf 63}: 115

\bibitem{wangzg2} Wang~Z~G, Liu~Z~C and Zhang~X~H, arXiv:0907.1467

\bibitem{x4350} Shen C P {\em et al.} (Belle Collaboration),
arXiv:0912.2383

\bibitem{zhangjr} Zhang~J~R and Huang~M~Q, arXiv:0905.4672

\bibitem{belle_z4430} Choi~S~K {\em et al.} (Belle Collaboration),
Phys. Rev. Lett., 2008, {\bf 100}: 142001

\bibitem{babar_z4430} Aubert~B {\em et al.} (BaBar Collaboration),
Phys. Rev. D, 2009, {\bf  79}: 112001

\bibitem{belle_z4430_dalitz}  Mizuk~R {\em et al.} (Belle Collaboration),
Phys. Rev. D, 2009, {\bf 80}: 031104

\bibitem{belle_z14050}  Mizuk~R  {\em et al.} (Belle Collaboration),
Phys. Rev. D, 2008, {\bf 78}: 072004

\bibitem{ktchao} Gao~Y~J, Zhang~Y~J, and Chao~K~T, hep-ph/0701009

\bibitem{y1s-decays} Shen C P {\em et al.} (Belle Collaboration), in preparation

\bibitem{belle2} SuperKEKB Task Force, KEK Report 2004-4.

\end{thebibliography}
\end{document}